\def\year{2015}
\documentclass[letterpaper]{article}
\usepackage{aaai}
\usepackage{times}
\usepackage{helvet}
\usepackage{courier}
\usepackage{graphicx}
\usepackage{microtype}
\nocopyright
\makeatletter
\renewcommand\@biblabel[1]{}
\makeatother

\begin{document}
%
\title{On Optimizing Human-Machine Task Assignments}
\author{\footnotesize\rm
	\begin{tabular}{rp{5.8in}}
		\textbf{Organizers:\thanks{This project was created via a world-wide, crowdsourced
				research process initiated by UC Santa Cruz, Stanford University, and Cornell Tech.}}& 
		\mbox{Andreas Veit,}
		\mbox{Michael Wilber,}
		\mbox{Rajan Vaish,}
		\mbox{Serge Belongie,}
		\mbox{James Davis.}
		\\
		\textbf{Top researchers:}& 
		\mbox{Vishal Anand,}
		\mbox{Anshu Aviral,}
		\mbox{Prithvijit Chakrabarty,}
		\mbox{Yash Chandak,}
		\mbox{Sidharth Chaturvedi,}
		\mbox{Chinmaya Devaraj,}
		\mbox{Ankit Dhall,}
		\mbox{Utkarsh Dwivedi,}
		\mbox{Sanket Gupte,}
		\mbox{Sharath N. Sridhar,}
		\mbox{Karthik Paga,}
		\mbox{Anuj Pahuja,}
		\mbox{Aditya Raisinghani,}
		\mbox{Ayush Sharma,}
		\mbox{Shweta Sharma,}
		\mbox{Darpana Sinha,}
		\mbox{Nisarg Thakkar,}
		\mbox{K.\ Bala Vignesh,}
		\mbox{Utkarsh Verma,}
		\\
		\textbf{Researchers:}& 
		\mbox{Kanniganti Abhishek,}
		\mbox{Amod Agrawal,}
		\mbox{Arya Aishwarya,}
		\mbox{Aurgho Bhattacharjee,}
		\mbox{Sarveshwaran Dhanasekar,}
		\mbox{Venkata Karthik Gullapalli,}
		\mbox{Shuchita Gupta,}
		\mbox{Chandana G,}
		\mbox{Kinjal Jain,}
		\mbox{Simran Kapur,}
		\mbox{Meghana Kasula,}
		\mbox{Shashi Kumar,}
		\mbox{Parth Kundaliya,}
		\mbox{Utkarsh Mathur,}
		\mbox{Alankrit Mishra,}
		\mbox{Aayush Mudgal,}
		\mbox{Aditya Nadimpalli,}
		\mbox{Munakala Sree Nihit,}
		\mbox{Akanksha Periwal,}
		\mbox{Ayush Sagar,}
		\mbox{Ayush Shah,}
		\mbox{Vikas Sharma,}
		\mbox{Yashovardhan Sharma,}
		\mbox{Faizal Siddiqui,}
		\mbox{Virender Singh,}
		\mbox{Abhinav S.,}
		\mbox{Pradyumna Tambwekar,}
		\mbox{Rashida Taskin,}
		\mbox{Ankit Tripathi,}
		\mbox{Anurag. D. Yadav}
	\end{tabular}
}
\maketitle
\begin{abstract}
When crowdsourcing systems are used in combination with machine inference systems in the real world, they benefit the most when the machine system is deeply integrated with the crowd workers. However, if researchers wish to integrate the crowd with ``off-the-shelf'' machine classifiers, this deep integration is not always possible.
This work explores two strategies to increase accuracy and decrease cost under this setting.
First, we show that reordering tasks presented to the human can create a significant accuracy improvement. Further, we show that greedily choosing parameters to maximize \emph{machine} accuracy is sub-optimal, and joint optimization of the \emph{combined} system improves performance.
\end{abstract}

\section{Introduction}

When crowdsourcing systems are deployed in the real world, the goal is often to maximize accuracy at a fixed price point or to minimize cost at a certain accuracy requirement. 
The best way to do this is by tightly integrating the machine and crowd worker within the overall end-to-end pipeline. For instance, the machine computation might use worker annotations as a prior to influence its results, or tasks for workers might be chosen and ordered adaptively using a Markov Decision Process~\cite{RussakovskyCVPR15}.

However, this tight integration is not always possible. Many real systems only provide outputs and cannot be heavily modified.
In these cases, the use of crowd workers is often restricted to a post-process that attempts to correct errors in the machine computation. In this scenerio, what kinds of strategies can maximize accuracy while minimizing costs?

To explore this question, we choose a representative task within the domain of computer vision: localizing objects in a large dataset. The goal is to detect all instances of certain objects of interest in the dataset. Machine systems can take images as input and automatically generate bounding boxes around objects of interest. Internal to the machine algorithm, to classify a potential detection as an object of interest or not, the algorithm employs a \emph{detection threshold} such that only detections with confidence scores above the threshold are returned.
Finding many correct objects implies also detecting many false positives. Because the detection threshold determines this tradeoff, it is often treated as the primary tunable parameter of machine vision algorithms.
The returned detections are then given to human workers, who we employ to remove false detections.
For our experiments, we adopt the classic UIUC-Cars dataset~\cite{agarwal2004learning}. As detector, we use Support Vector Machines trained on Histograms of Ordered Gradients as a representative ``out-of-the-box'' machine vision system.

Our objective is to maximize the overall accuracy of the machine-crowd pipeline on the dataset given a certain time budget. We vary the time budget by presenting the humans with only a fraction of all detections. If humans look at a large fraction of detections the accuracy improvement will be large, however the average time cost per image in the dataset will also be large. If humans look at only a few images, the average accuracy of the entire dataset will show little improvement, but the time cost will be low. We plot the tradeoff between cost and accuracy as a curve.

\begin{figure}[t]
	\includegraphics[width=0.95\linewidth]{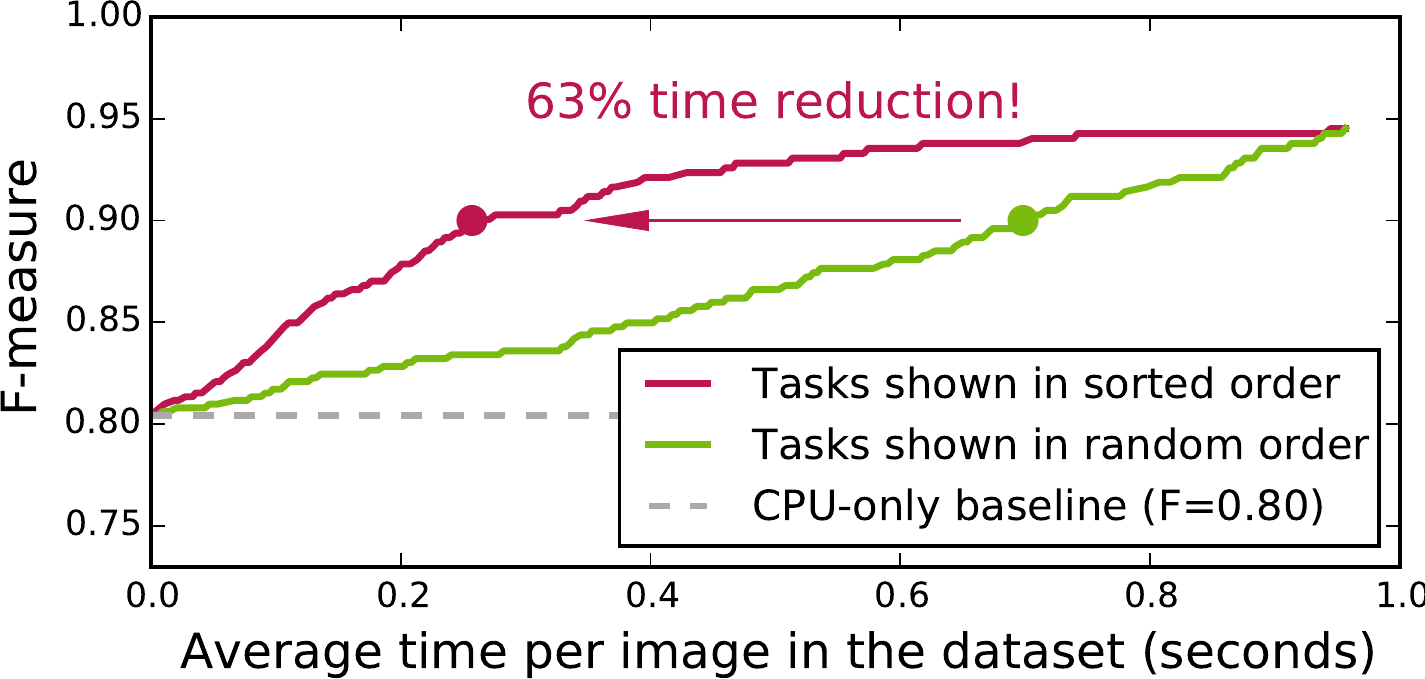}
	\caption{\footnotesize
		Consider a simple localization task where crowd workers refine the output of a machine classifier. At a threshold of 0.5, baseline accuracy starts at 0.80 (gray dotted line). If we show random tasks to human workers, accuracy improves (green), but if we order tasks by increasing machine confidence (purple), we can reduce the time requirement dramatically at a given target accuracy.
	}
	\label{fig:reorder}
\end{figure}
The primary contribution of this work is a description and analysis of two strategies for improving the cost-accuracy curve. In Task Ordering we consider the impact of using the machine vision algorithm's confidence score as a way to order human tasks. In Joint Optimization we consider how changing the machine threshold parameter impacts results.

\section{Task Ordering}\label{sec:reorder-images}
First we analyze the impact of the ordering in which we present tasks to human workers. The machine algorithm processes each image and returns detections that are above a pre-defined threshold.  We then show the detections to workers until the time budget runs out. If all detections are shown to workers, then the ordering is unimportant. However, under a limited time budget we can present only a fraction of the images to the workers. Consequently, the ordering of tasks is relevant, as it affects which images are presented and corrected by the crowd.

Does it matter which detections are given to the crowd first? Clearly it does. Each true detection presented to the crowd is costing time, but not impacting accuracy. Each false detection which is identified and removed by the crowd improves accuracy. We seek an oracle which will allow us to present only false detections to the crowd. Unfortunately we necessarily lack such an oracle, because if it existed, we could use it to modify the machine algorithm to remove the false detections in the first place.

In our experiments, we consider two different task orderings for the human workers: First, as a baseline, we present the detections in random order. Second, we order the detections such that those with lowest machine confidence are presented to the humans first. The motivation behind this ordering is that we suspect the machine of making more mistakes among detections that are closer to the detection threshold. This is equivalent to assuming that the machine algorithm confidence score can be used as an approximate oracle for predicting false detections.  

The results of our trials are depicted in Figure~\ref{fig:reorder}.
We plot the accuracy of the entire pipeline versus the amount of human worker involvement.
We observe that re-ordering human tasks can greatly reduce the required human work. In particular, we can see that to achieve an accuracy, in terms of F-measure, of $0.90$, reordering the tasks based on confidence scores can reduce the required human time cost by $62.9\%$ compared to random ordering. 

\section{Joint Optimization}
We investigate the proper choice of the machine detection threshold parameter when also utilizing human labor. The machine returns bounding boxes around potential detections with accuracy dependent on this threshold parameter. Typically, the detection threshold is tuned on a validation dataset so as to maximize accuracy of the machine detections. One approach for combining machine and human computation would be to use the threshold that leads to the highest machine accuracy, and subsequently use human workers to improve the reported detections. However, the choice of the threshold also changes which detections are presented to the human workers. As shown in the previous section, this can greatly influence the accuracy of the overall pipeline.

\begin{figure}[t]
	\includegraphics[width=1\linewidth]{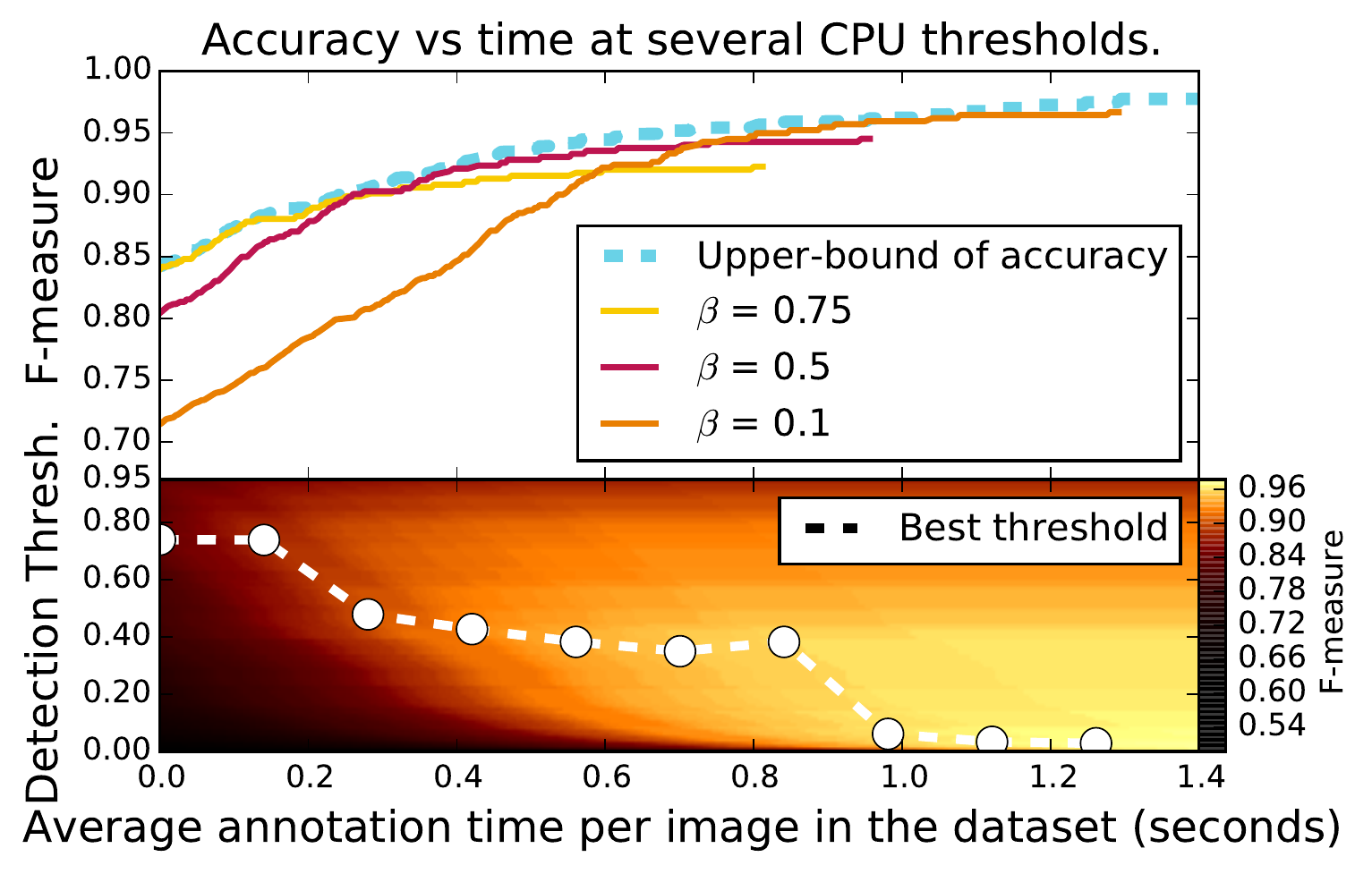}
	\caption{\label{fig:oracle}\footnotesize
		Performance is dependent on the detection threshold because it determines which images will be shown to crowd workers. Top: Plots F-measure given three different thresholds (Purple line is the same as in Fig.~\ref{fig:reorder}). Note that no single threshold is best. The blue line is the best performance at the best possible threshold for that specific time budget. Bottom: Plots F-measure given a certain time budget and a certain threshold. The best threshold for each time budget is highlighted in white. Note that the optimum threshold for machine-only computation (0.0 time budget) is no longer optimum when a human labor time budget is available.
	}
\end{figure}

Should the parameters of the machine algorithm be set to their independently optimal values, or would some sub-optimal parameter choice lead to overall better performance of the joint system? The results of our experiments are shown in Figure~\ref{fig:oracle}. We observe that there is not a single optimal threshold to pick. The optimal threshold depends on the time budget and accuracy requirement. As the time budget increases it becomes desirable to choose a threshold that \textit{reduces} the initial accuracy of the machine algorithm. The upper bound on the achievable accuracy for each time budget is shown by the blue dashed line in the top part of Figure~\ref{fig:oracle}. The respective optimal threshold is shown in the lower part.

\section{Conclusion}
When human computation is used to improve the results of machine algorithms, deep integration of human labor into the algorithm is not always possible, and budgets are frequently limited. We investigate two strategies for optimizing the use of human labor in this context.

We find that Task Ordering can have a dramatic impact on the  human time budget required to achieve a particular target accuracy. A cost savings of 62\% was achieved in our experiments. We also find that it is insufficient to let domain experts optimize parameters and performance of machine algorithms, while human computation experts optimize the use of the crowd. Optimal performance was achieved by first finding the accuracy and budget requirements, and then setting the parameters of the machine algorithm to support the joint machine-human pipeline.

\bibliography{main}
\bibliographystyle{aaai}
\end{document}